# First assessment of the plant phenology index (PPI) for estimating gross primary productivity in African semi-arid ecosystems

A.M. Abdi[a, b, *], N. Boke-Olén[b], H. Jin[a], L. Eklundh[a], T. Tagesson[a], V. Lehsten[a, c], J. Ardö[a]

[a] Department of Physical Geography and Ecosystem Science, Lund University, Sölvegatan 12, SE-223 62 Lund, Sweden
[b] Centre for Environmental and Climate Research, Lund University, Sölvegatan 37, SE-223 62 Lund, Sweden
[c] Swiss Federal Institute for Forest, Snow and Landscape research (WSL), Zürcherstr. 11, CH-8903 Birmensdorf, Switzerland

## ARTICLE INFO

## ABSTRACT

The importance of semi-arid ecosystems in the global carbon cycle as sinks for $CO_2$ emissions has recently been highlighted. Africa is a carbon sink and nearly half its area comprises arid and semi-arid ecosystems. However, there are uncertainties regarding $CO_2$ fluxes for semi-arid ecosystems in Africa, particularly savannas and dry tropical woodlands. In order to improve on existing remote-sensing based methods for estimating carbon uptake across semi-arid Africa we applied and tested the recently developed plant phenology index (PPI). We developed a PPI-based model estimating gross primary productivity (GPP) that accounts for canopy water stress, and compared it against three other Earth observation-based GPP models: the temperature and greenness (T-G) model, the greenness and radiation (G–R) model and a light use efficiency model (MOD17). The models were evaluated against in situ data from four semi-arid sites in Africa with varying tree canopy cover (3–65%). Evaluation results from the four GPP models showed reasonable agreement with in situ GPP measured from eddy covariance flux towers (EC GPP) based on coefficient of variation ($R^2$), root-mean-square error (RMSE), and Bayesian information criterion (BIC). The G–R model produced $R^2 = 0.73$, RMSE = 1.45 g C m$^{-2}$ d$^{-1}$, and BIC = 678; the T-G model produced $R^2 = 0.68$, RMSE = 1.57 g C m$^{-2}$ d$^{-1}$; the MOD17 model produced $R^2 = 0.49$, RMSE = 1.98 g C m$^{-2}$ d$^{-1}$, and BIC = 800. The PPI-based GPP model was able to capture the magnitude of EC GPP better than the other tested models ($R^2 = 0.77$, RMSE = 1.32 g C m$^{-2}$ d$^{-1}$, and BIC = 631). These results show that a PPI-based GPP model is a promising tool for the estimation of GPP in the semi-arid ecosystems of Africa.

## 1. Introduction

The uptake of $CO_2$ by terrestrial vegetation through photosynthesis, termed gross primary productivity (GPP), is the largest flux of the global carbon cycle (Le Quere et al., 2009). Globally, GPP accounts for around 120 Pg of carbon (PgC) and its spatiotemporal variability is not fully understood due to complex processes related to plant physiology and environmental controls on photosynthesis (Xia et al., 2015). Several modeling approaches attempted to describe the process whereby carbon is assimilated by plants through photosynthesis (Anav et al., 2015; McCallum et al., 2009). Vegetation indices derived from remote sensing have been used to estimate GPP empirically and upscale it at regional or continental scales (Sjöström et al., 2011, 2009). Two commonly used empirical models based on remotely-sensed enhanced vegetation index (EVI) are the temperature-greenness (T-G) model (Sims et al., 2008) and the greenness-radiation (G–R) model (Gitelson et al., 2006). The popularity of these models lies in their simplicity and the fact that they are not dependent on site measurements. Another popular approach, first described by Monteith (1972), is based on the efficiency with which absorbed solar radiation is converted into photosynthesized carbon (light use efficiency, LUE). Presently, the MODerate resolution Imaging Spectroradiometer (MODIS) onboard the Terra and Aqua satellites is the only Earth-observation (EO) platform that provides near-real-time estimates of terrestrial carbon uptake using the LUE method. At the core is the MOD17 algorithm (Running et al., 2004) that uses MODIS spectral data and climatic drivers in an LUE model (Heinsch et al., 2003).

In a recent review, Tang et al. (2016) expressed concern whether spectral vegetation indices can really capture the primary productivity cycle since they represent the endpoint of plant physiological processes such as assimilation, respiration and phytomass accretion. These concerns underscore the notion that vegetation indices are not reduced to zero during periods of drought and thus do not explicitly show when photosynthesis ends (Walther et al., 2016). On the other hand, actual LUE has large spatial and temporal heterogeneity that may not be captured at regional or continental scales. Furthermore, the lack of high-resolution gridded meteorological data limits the derivation of environmental scalars such as temperature (Turner et al., 2003; Zhao et al., 2005). Primary productivity estimates provided by MODIS may not be representative of field conditions as demonstrated by Sjöström et al. (2013). They found that the MOD17 algorithm underestimated maximum LUE by an average of 1.2 g C MJ$^{-1}$ for most of the sites in their study.

* Corresponding author at: Department of Physical Geography and Ecosystem Science, Lund University, Sölvegatan 12, SE-223 62 Lund, Sweden.
  Email address: hakim.abdi@gmail.com (A.M. Abdi)

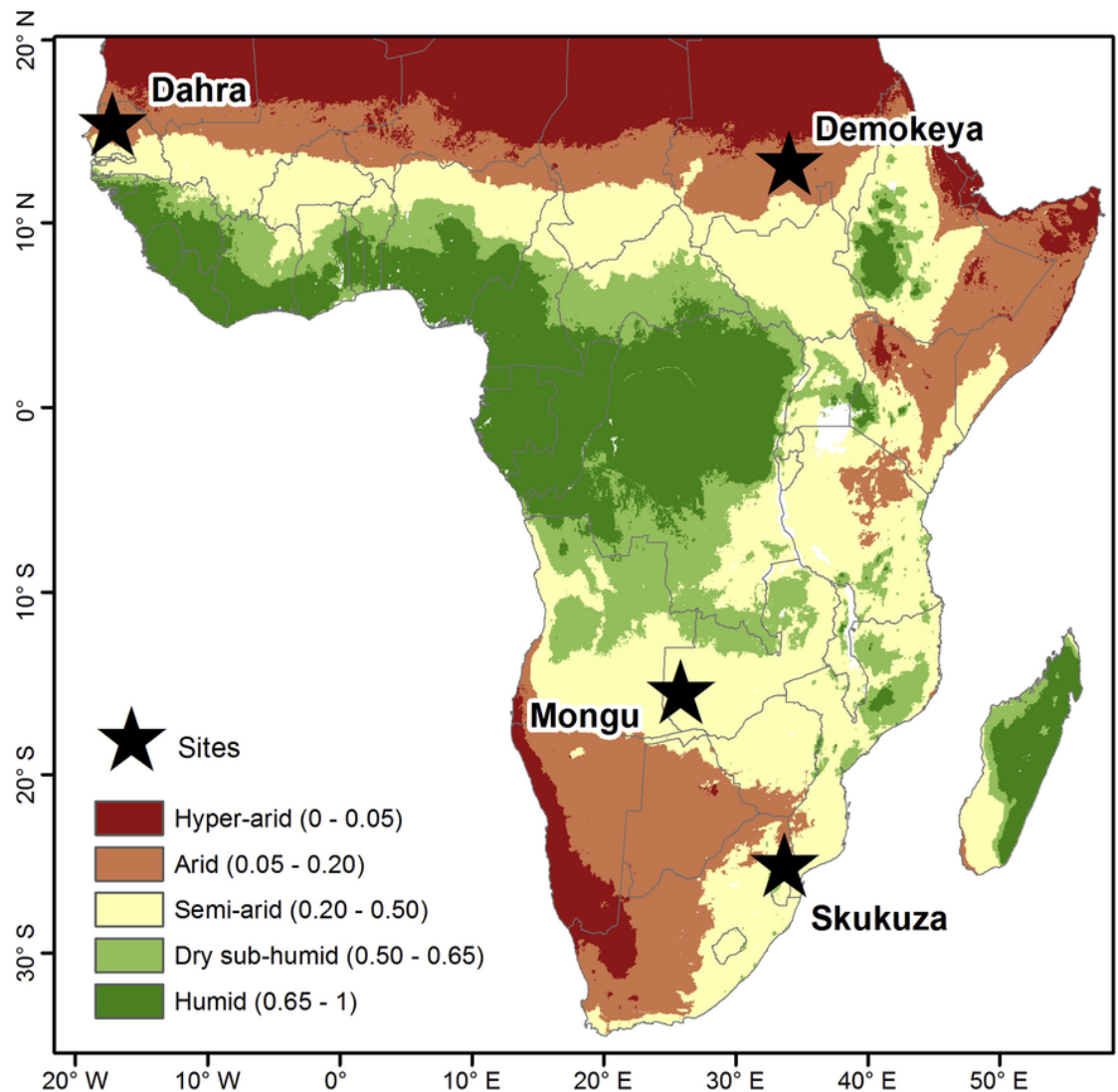

Fig. 1. Location of the sites and the aridity zones in the study region.

Considering these shortcomings, Jin and Eklundh (2014) developed a physically-based vegetation index in an effort to achieve a higher consistency of EO-based vegetation indices that better explain the relationship between canopy development and GPP. This index, called the plant phenology index (PPI), is near-linear to green leaf area of the vegetation canopy and thus may be able to better represent carbon assimilation. It has proven superior to other established indices for estimating spring phenology over the northern hemisphere (Jin et al., 2017; Karkauskaite et al., 2017).

The importance of semi-arid regions in the global carbon cycle has recently been highlighted by Ahlström et al. (2015). Excluding the hyper-arid Sahara desert, most of Africa comprises arid and semi-arid zones (Abdi et al., 2016). These zones are classified as grassland (39%), savanna (23%) or woody savanna (5%) according to the MODIS land cover product (Sulla-Menashe and Friedl, 2015). Around 11% of sub-Saharan Africa is classified as tropical forest. Africa is a carbon sink of around 0.6 Pg C yr$^{-1}$ (Keenan and Williams, 2018), but there are gaps in our knowledge, particularly about the role of savannas and dry tropical woodlands (Valentini et al., 2014). Observational carbon cycle studies are underrepresented in Africa relative to other regions (Ciais et al., 2011; Valentini et al., 2014). For example, of the six eddy covariance flux tower sites in Africa that are presently freely available in the FLUXNET2015 dataset (http:/fluxnet.fluxdata.org/data/fluxnet2015-dataset/), only four are in arid or semi-arid ecosystems. It is, therefore, important to test new observational data over Africa that have the potential to be extended across large scales. The core objective of this study is to derive an empirical GPP model based on the recently developed PPI and evaluate its performance against other GPP models without the use of ground-based meteorological data. This will be the first derivation and assessment of a PPI-based model that estimates GPP in African semi-arid ecosystems.

**Table 1**
Descriptions and physical characteristics of the sites included in this study.

| Site, Country (Lon, Lat) | Ecosystem Type (Tree Cover, %) | MAP (mm) | MAT (°C) | Measurement Years | Site References |
|---|---|---|---|---|---|
| Demokeya, Sudan (30.47, 13.28) | Sparse acacia savanna (7%) | 320 | 30 | 2007 – 2009 | Ardö et al. (2005-2009); Ardö et al. (2008) |
| Dahra, Senegal (-15.43, 15.40) | Open woody savanna (3%) | 416 | 29 | 2010 – 2013 | Tagesson et al. (2010-2013); Tagesson et al. (2015) |
| Skukuza, South Africa (31.49, -25.01) | Wooded grassland (30%) | 547 | 22 | 2001 – 2005 | Archibald et al. (2009); Scholes (2000-2013) |
| Mongu, Zambia (23.25, -15.43) | Miombo woodland (65%) | 879 | 25 | 2008 – 2009 | Kutsch et al. (2000-2009); Merbold et al. (2009) |

**Table 2**
In situ measurements extracted from the FLUXNET2015 dataset for each of the four sites.

| Variables | Description | Units |
|---|---|---|
| GPP_DT_CUT_MEAN | Gross primary production using the daytime partitioning method from Lasslop et al. (2010), average of GPP versions. | g C m$^{-2}$ d$^{-1}$ |
| VPD_F | Vapor pressure deficit consolidated from VPD_F_MDS and VPD_ERA methods | hPa |
| SWC_F_MDS_1 | Volumetric soil water content of the upper layer, gap-filled with MDS | % |
| SW_IN_POT | Potential incoming shortwave radiation (top of atmosphere) | W m$^{-2}$ |

## 2. Data and methods

### 2.1. Study sites

This study focused on four eddy covariance flux tower sites, two in the Sahel and two in southern Africa (Fig. 1). These sites were selected because they were: (1) in either arid or semi-arid ecosystems in Africa, (2) they were harmonized and standardized, thus reducing data processing-related errors, (3) they were easily accessible and freely available, and (4) they had at least two years of reasonably continuous data (i.e. no long periods of no-data values). The sites are located in a climatological gradient with a mean annual precipitation (MAP) ranging between 320–879 mm and mean annual temperature (MAT) between 22–30 °C and are generally representative of arid and semi-arid savanna ecosystems in Africa (Table 1).

### 2.2. Eddy covariance flux tower data

The field data are part of the FLUXNET2015 dataset which is a harmonized and standardized global dataset of micrometeorological, energy, and net ecosystem exchange of $CO_2$ between the atmosphere and terrestrial biosphere (FLUXNET, 2015). The data processing pipeline in the FLUXNET2015 dataset ensures inter-comparison and quality assurance and control across multiple sites (Vuichard and Papale, 2015).

The four eddy covariance flux tower sites (Table 1) consisted of 14 site-years and covered two main ecosystem types: savanna (SAV) and woody savanna (WSA). Daily measurements for four variables were extracted for each site (Table 2) and averaged into 8-day intervals. Variables were screened using quality flags so that only samples that are either measured (flag = 0) or good quality (flag = 1) were retained. These data are all Tier-1 level under the fair-use data policy of FLUXNET, meaning that the data are open and free for scientific purposes (http://fluxnet.fluxdata.org/data/data-policy/).

### 2.3. Earth observation data and models

#### 2.3.1. The plant phenology index

Leaf area index (LAI), the one-sided area of leaf per unit area of ground surface (m$^2$ m$^{-2}$), represents the total leaf surface area that intercepts incoming radiation. It denotes leaf area that is potentially available for the exchange of gases between the atmosphere and vegetated land surfaces (Cowling and Field, 2003). Depending on the phenological stage, LAI can be split into photosynthetic and non-photosynthetic portions. Photosynthetic (green) LAI is closely related to canopy chlorophyll content (Ciganda et al., 2008) while non-photosynthetic (brown) LAI is linked to dry or senescing vegetation (Pinter et al., 1983). PPI is nearly linear to green LAI and was expressed by Jin and Eklundh (2014) as:

$$PPI = -K \times \ln \left( \frac{(NIR - RED)_{max} - (NIR - RED)}{(NIR - RED)_{max} - (NIR - RED)_{soil}} \right) \quad (1)$$

where PPI is in LAI units (m$^2$ m$^{-2}$); K is a gain factor that is dependent on sun zenith angle, geometry of leaf angular distribution, and the instantaneous diffuse fraction of solar radiation (Jin and Eklundh, 2014); NIR and RED are sun-sensor geometry-corrected near infrared and red bands, respectively; the subscript "max" indicates per-pixel canopy maximum calculated from satellite observations covering 2000–2014; the subscript "*soil*" is soil reflectance, which was derived from ASTER Spectral Library (Baldridge et al., 2009) and comprises 41 reflectance samples representing major global soil types. See Jin and Eklundh (2014) for a detailed description of PPI.

PPI was downloaded from the DataGURU open access repository of Lund University (https://dataguru.lu.se/). This is the initial version of the index developed by Jin and Eklundh (2014) based on Collection 5.1 (C5.1) MODIS spectral data and is arranged under the MODIS tiling system. We downloaded four tiles corresponding to the location of each site. PPI values were averaged over approximately 3 km x 3 km around the coordinates of each flux tower to facilitate comparability with the MODIS data acquired from the Oak Ridge National Laboratory Distributed Active Archive Center (ORNL DAAC, 2008).

#### 2.3.2. The PPI GPP model

Leaf chemical processes are likely to be influenced by the feedback between canopy, near-surface air temperature and relative humidity (Hashimoto et al., 2008). EO-derived land surface temperature (LST) correlates reasonably well with vapor pressure deficit (VPD), which has been used as down-regulating mechanism in modeling GPP (Hashimoto et al., 2008; Wu et al., 2010). VPD is the difference between the amount of moisture in the air and the amount of moisture the air could hold at saturation. According to Fick's first law of diffusion, particle fluxes move from regions of high concentration to regions of low concentration. High VPD indicates the atmosphere is low in moisture, which causes water to exit the plant during transpiration. Hence, plants may close their stomata as a physiological response to high VPD in order to prevent water loss (Meinzer et al., 1993). With the stomata closed, plants cannot absorb $CO_2$ and photosynthesis declines (Lambers et al., 2008; Slot and Winter, 2017). Here, VPD was derived from the MODIS daytime LST product (MOD11A2, C5.1) using Eq. 2. This was based on the observed relationship between MODIS daytime LST and in situ VPD from flux towers across all 14 site-years. The rationale for deriving VPD from LST was to facilitate a fully EO-based estimation of GPP.

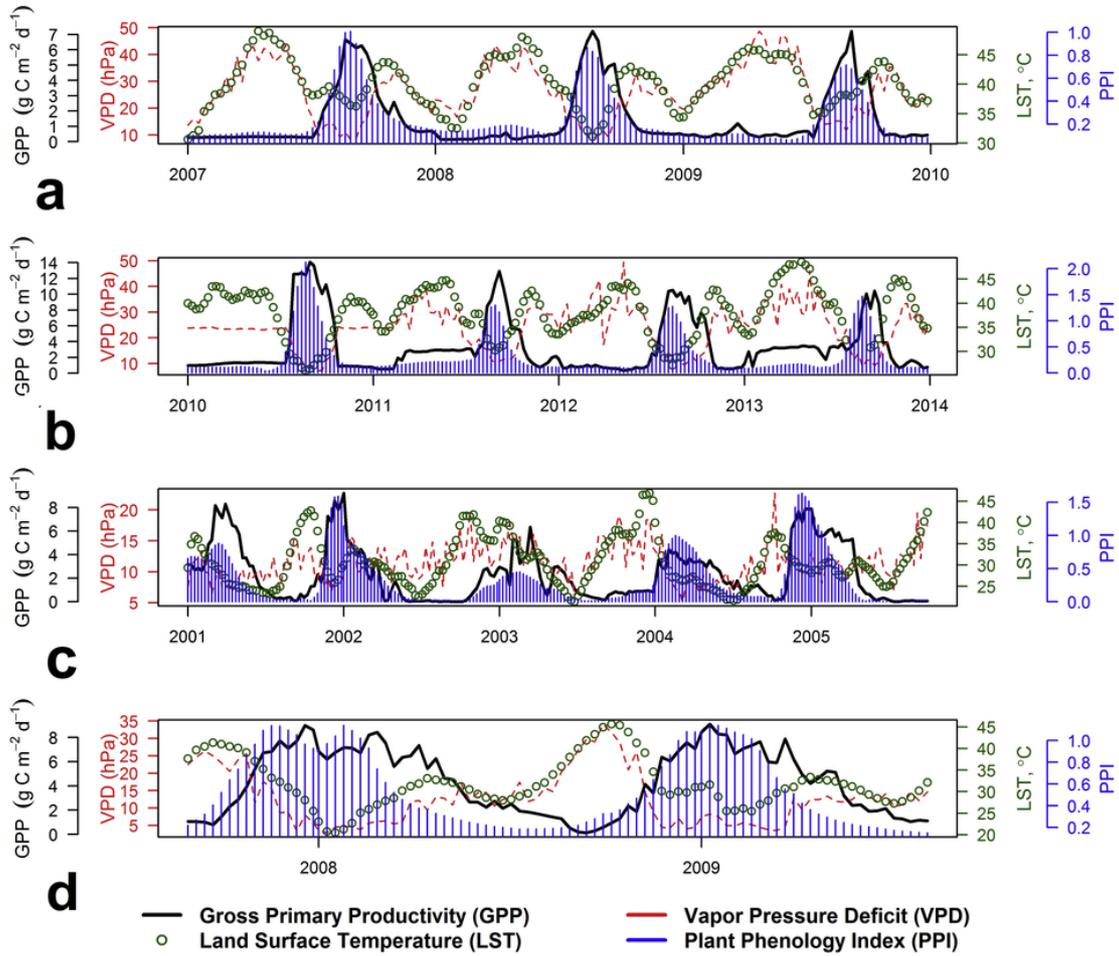

**Fig. 2.** Seasonality and inter-annual variability of 8-day averaged PPI and MOD11A2 LST, and the FLUXNET2015 derived EC GPP and VPD for **(a)** Demokeya, **(b)** Dahra, **(c)** Skukuza, **(d)** Mongu.

$$VPD_{LST} = -2.74 + (2.57 * 1.06^{LST_{Day}})$$ (2)

where $VPD_{LST}$ is vapor pressure deficit (hPa) derived from LST; $LST_{Day}$ (°C) is the 8-day MOD11A2 product. GPP in arid and semi-arid ecosystems has been shown to decline when VPD exceeds 20 hPa (Abdi et al., 2017; Rezende et al., 2016; Sjöström et al., 2013; Whitley et al., 2011; Zhang et al., 2007). Mean in situ GPP across all four sites was $3.87 \pm 3.15$ g C m$^{-2}$ d$^{-1}$ at VPD values less than 20 hPa, and at VPD values greater than 20 hPa mean GPP was $1.21 \pm 1.04$ g C m$^{-2}$ d$^{-1}$. Stomatal conductance was not measured in this study, however, it is likely that increase in VPD can cause stomatal closure (Meinzer et al., 1993), thereby reducing GPP. Therefore, we used $VPD_{LST}$ as a down-regulating scalar on GPP by scaling $VPD_{LST}$ so that GPP declines with increasing $VPD_{LST}$.

$$VPD_{Scaled} = 1 - \left[ \frac{VPD_{LST} - \min(VPD_{LST})}{\max(VPD_{LST}) - \min(VPD_{LST})} \right]$$ (3)

where $VPD_{scaled}$ is a defined as an inversely normalized function of $VPD_{LST}$.

The calibration procedure influences model accuracy as it is when the relationship between GPP and PPI is established and is detailed in

Section 2.4. Briefly, parameterization was performed using 1000 bootstrapped permutations of ordinary least-squares regression between PPI and EC GPP. Bootstrapping is a procedure in which the calibration dataset is randomly sampled *n* times with replacement to empirically estimate the sampling distribution (Efron, 1979). This results in the regression statistics that define the PPI – EC GPP relationship. Then, GPP was defined as the product of parameterized PPI and $VPD_{Scaled}$:

$$PPI\ GPP = PPI_{parameterized} \times VPD_{Scaled} \times m$$ (4)

where m is the slope coefficient of the parameterization of PPI and has the units of g C m$^{-2}$ d$^{-1}$. The underlying assumption of the model depicted in Eq. 4 is that green LAI is directly linked to GPP as a proxy for canopy chlorophyll content, and that GPP is reduced by an increase in VPD.

### 2.3.3. MODIS GPP

LUE is based on the premise that GPP is proportional to the product of photosynthetically active radiation (PAR), the fraction of PAR that is absorbed by the vegetation canopy (fPAR), and a parameter that translates absorbed energy into assimilated carbon. Production efficiency models such as MOD17 have to include ingenious workarounds to account for LUE, which is not directly measured. The adoption of a maximum LUE ($\varepsilon_{max}$) parameter is one such workaround. This parameter is defined as the hypothetical biome-specific ideal condition (Monteith, 1972). MOD17 utilizes a Biome Properties Look-Up Table (BPLUT) that defines $\varepsilon_{max}$ for 11 general plant functional types (PFT) based on the assumption that $\varepsilon_{max}$ varies little within a particular PFT (Monteith and

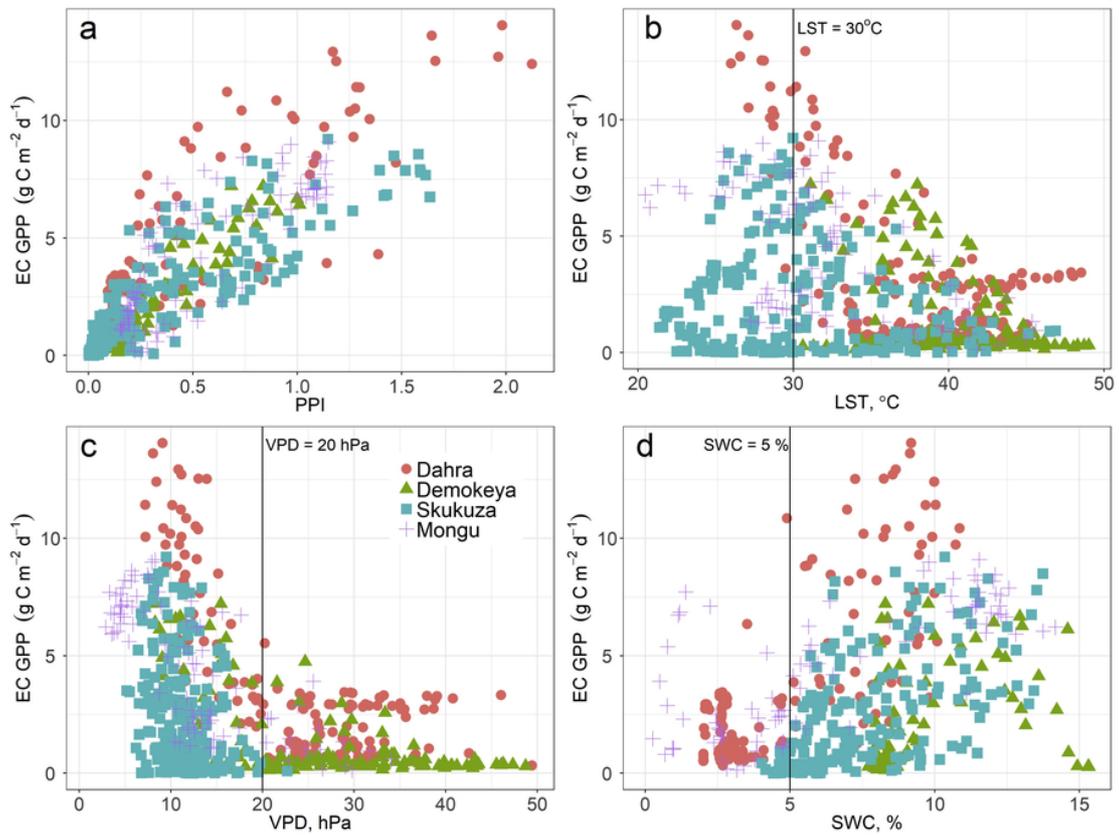

**Fig. 3.** Scatterplots of individual site-based relationships between 8-day EC GPP and LST, VPD and SWC.

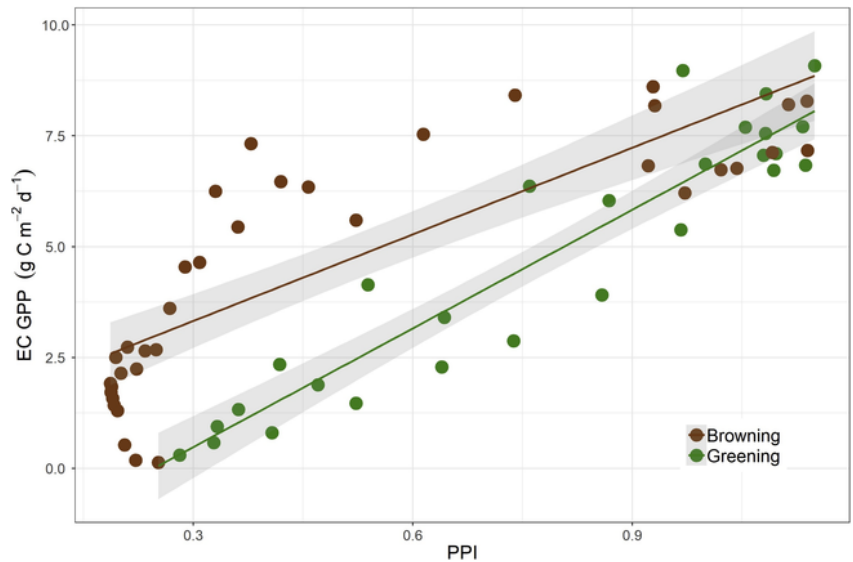

**Fig. 4.** Close-up of the seasonal hysteresis in the relationship between EC GPP and PPI at Mongu. The gray shading indicates the 95% confidence interval.

Moss, 1977).

$$GPP_{MOD17} = fPAR \times PAR \times T_{min} \times VPD \times \varepsilon_{max} \qquad (5)$$

where $T_{min}$ is minimum temperature that reduces photosynthesis in colder climates; VPD is vapor pressure deficit, which reduces photosynthesis when temperature is high and relative humidity is low. The MOD17 8-day GPP product (MOD17A2, C5.1) within a 3 km x 3 km area around each flux tower site was downloaded for each of the four sites from ORNL DAAC.

### 2.3.4. T-G GPP model

The temperature and greenness model (Sims et al., 2008) uses the product of scaled LST and the enhanced vegetation index (EVI) to estimate GPP. It is based on the idea that GPP has a generally strong correlation with EVI, and that LST accounts for temperature controls on GPP.

$$GPP_{T-G} = EVI_{scaled} \times LST_{scaled} \times m \qquad (6)$$

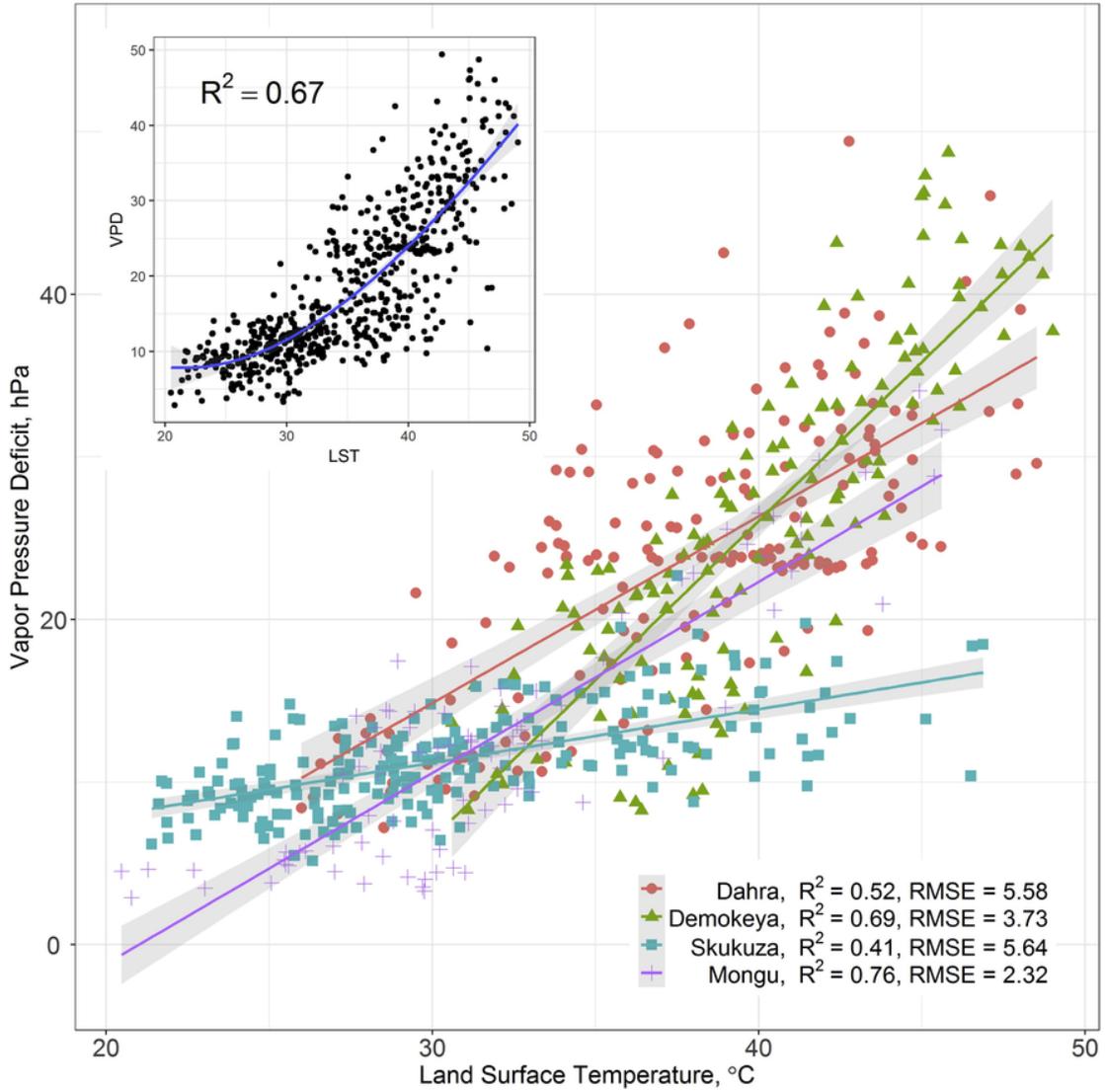

**Fig. 5.** Individual and all-site (inset) relationships between 8-day in situ VPD derived from flux towers and 8-day daytime *LST* from the MOD11A2 product. The solid line is the regression line with 95% confidence intervals in shaded gray.

**Table 3**
Summary of regression statistics resulting from the bootstrapping of EC GPP and PPI across the four sites. $\beta_0$ is the intercept, $\beta_1$ is the slope, $R^2$ is the coefficient of determination, F is the F-statistic, DF is the degrees of freedom, SE is the standard error, and RMSE is the root-mean-square error in g C m$^{-2}$ d$^{-1}$, P is the probability value, and BIC is the Bayesian information criterion.

| Response: EC GPP | | $\beta_0$ | $\beta_1$ | $R^2$ | F | DF | SE | RMSE | P | BIC |
|---|---|---|---|---|---|---|---|---|---|---|
| PPI | Calibration | 0.516 | 6.166 | 0.77 | 1354 | 468 | 0.175 | 1.45 | <0.001 | 1707 |
| | Evaluation | 0.017 | 0.999 | 0.75 | 571 | 185 | 0.042 | 1.41 | <0.001 | 673 |

where m is the slope coefficient of the parameterization of the T-G model and EC GPP, and has the units of g C m$^{-2}$ d$^{-1}$.

EVI$_{scaled}$ reduces GPP to zero when EVI $\approx$ 0.1 (Eq. 7). LST$_{scaled}$ reduces to 1 when LST = 30 °C, and to 0 when LST = 0 or when it reaches 50 °C (Eq. 8).

$$EVI_{scaled} = EVI - 0.1 \tag{7}$$

$$LST_{scaled} = \min\left[\left(\frac{LST}{30}\right); (2.5 - (0.05 \times LST))\right] \tag{8}$$

### 2.3.5. G–R GPP model

The greenness and radiation model (Gitelson et al., 2006) is based on the idea that total chlorophyll content of a canopy is the primary factor influencing the amount of PAR absorbed by green vegetation. Chlorophyll is an important pigment for absorbing PAR and is crucial for plant productivity (Whittaker and Marks, 1975). The G–R model was originally formulated GPP as the product of total canopy chlorophyll (Chl) and incoming PAR at the top of the canopy (PAR$_{in}$):

$$GPP_{G-R} = Chl \times PAR_{in} \tag{9}$$

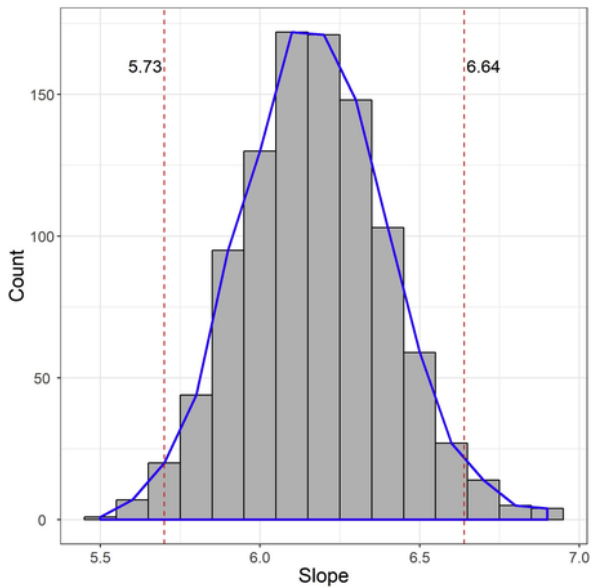

**Fig. 6.** Bootstrap distribution histogram and density curve of the EC GPP – PPI slope based on 1000 replicates (bias = 0.005). The dashed lines represent the 2.5% ($\beta_1$ = 5.73) and 97.5% ($\beta_1$ = 6.64) confidence intervals.

In Eq. 9, quantification of Chl was done by measuring in situ bi-weekly reflectance, green and total LAI, and PAR$_{in}$ was measured using point quantum sensors (Gitelson et al., 2006). The model was later extended by substituting in situ Chl with a remotely sensed vegetation index that is closely related to chlorophyll content such as EVI. PAR$_{in}$ was replaced with potential (top of atmosphere) incoming PAR (PAR$_{pot}$) to eliminate the dependence on in situ measurements and reduce fluctuations present in PAR$_{in}$ that introduce uncertainty and noise to the model (Gitelson et al., 2012; Peng et al., 2013):

$$GPP_{G-R} = EVI \times PAR_{pot} \times m \tag{10}$$

where m is the slope coefficient of the parameterization of the G–R model and EC GPP, and has the units of g C m$^{-2}$ d$^{-1}$. PAR$_{pot}$ is calculated as 40% of potential incoming shortwave radiation at the top of the atmosphere (SW_IN_POT, Table 2). PAR$_{pot}$ can also be independently calculated using the method described in Monteith and Unsworth (2013).

### 2.4. Model calibration and statistical analysis

All the EO data were smoothed with a Savitzky-Golay filter (Savitzky and Golay, 1964) in the software package TIMESAT (Jönsson and Eklundh, 2004) using a window size of 11 points, and assuming a single growing season. The remaining statistical analysis and visualization were done in R 3.4.3 (R Core Team, 2017). In order to assess the model performance and provide an independent evaluation, the data were split into two portions based on the year of measurement. This was done by randomly assigning 70% of the dataset (10 site-years, ˜463 samples) for calibration and using the remaining 30% (4 site-years, ˜184 samples) for evaluation. This combination was permuted in 1000 bootstrap replicates in order to get robust estimates of the regression coefficients. We used an ordinary least-squares regression to compare the GPP models against site-based GPP acquired from eddy covariance flux towers. We used the coefficient of determination ($R^2$) to measure variance explained by each model and the root-mean-square error (RMSE, Eq. 11) to assess the absolute fit of models to the data.

$$RMSE = \sqrt{\sum_{i=1}^{n} \frac{(OBS - PRED)^2}{n}} \tag{11}$$

where *OBS* is EC GPP (g C m$^{-2}$ d$^{-1}$), *PRED* is model-estimated GPP (g C m$^{-2}$ d$^{-1}$), and *n* is the number of sample points.

We also computed the Bayesian information criterion (BIC, Eq. 12) in order to account for the variations in sample sizes in the models (Burnham and Anderson, 2004; Schwarz, 1978). Differences in sample size between the different models were caused by a combination of some years having slightly more data than others due to missing values and the random splitting of site-years into calibration and evaluation subsets. BIC estimates the posterior probability of a model based on the log of the marginal likelihood for data with an uninformative prior so that the lower the BIC, the close a model is to being "true".

$$BIC = -2\ln(L) + q\ln(n) \tag{12}$$

where L is the maximum of the likelihood function, q is the parameter number, and n is the sample size. BIC levies a penalty term $q\ln(n)$ on increasing parameters and sample sizes to prevent over-fitting.

## 3. Results and discussion

### 3.1. Inter-annual and seasonal variations in the eddy covariance-derived variables

VPD seasonality corresponded well with seasonal and inter-annual pattern of GPP (Fig. 2). LST seasonality matched that of VPD at all sites except at the southernmost site, Skukuza (Fig. 2c), where VPD variability was high. The quick decrease in carbon assimilation at the Sahelian sites at the start of the dry season is probably due to the senescence of the herbaceous layer and the limited assimilation during the dry season is due to the sparse tree cover that can access deeper water (Ardö et al., 2008). A more gradual decrease in EC GPP was observed at the woodland site, Mongu, due to the higher density of trees at this site. PPI followed the seasonal progression of EC GPP generally well across all sites, however the green-up and brown-down phases had a distinct offset that contributed to seasonal hysteresis (discussed in the next section). Maximum GPP during the growing season was between 7.5–9.5 g C m$^{-2}$ d$^{-1}$ at most sites except Dahra, which had a maximum of 14 g C m$^{-2}$ d$^{-1}$, despite having considerably less tree canopy cover (TCC = ˜3%) than either Skukuza (30%) or Mongu (65%). The high GPP at Dahra has been attributed to a number of possible factors, including grazing-induced compensatory growth (Tagesson et al., 2016a) and increase in relative humidity during the growing season (Steiner et al., 2009; Tagesson et al., 2016b).

**Table 4**
Summary of the analyses between EC GPP and each of the four GPP models across all sites based on 8-day time series. $\beta_0$ is the intercept, $\beta_1$ is the slope, $R^2$ is the coefficient of determination, F is the F-statistic, DF is the degrees of freedom, SE is the standard error, and RMSE is the root-mean-square error in g C m$^{-2}$ d$^{-1}$, P is the probability value, and BIC is the Bayesian information criterion.

| Response: EC GPP | | $\beta_0$ | $\beta_1$ | $R^2$ | F | DF | SE | RMSE | P | BIC |
|---|---|---|---|---|---|---|---|---|---|---|
| PPI GPP | Calibration | 0.006 | 1.000 | 0.76 | 1505 | 456 | 0.026 | 1.38 | <0.001 | 1615 |
| | Evaluation | 0.004 | 1.011 | 0.77 | 677 | 180 | 0.039 | 1.32 | <0.001 | 631 |
| G-R Model | Calibration | 0.005 | 1.006 | 0.72 | 1211 | 463 | 0.029 | 1.51 | <0.001 | 1723 |
| | Evaluation | −0.038 | 1.029 | 0.73 | 524 | 183 | 0.045 | 1.45 | <0.001 | 678 |
| T-G Model | Calibration | −0.015 | 1.007 | 0.65 | 902 | 459 | 0.033 | 1.67 | <0.001 | 1795 |
| | Evaluation | −0.115 | 1.061 | 0.68 | 434 | 183 | 0.052 | 1.57 | <0.001 | 707 |
| MOD17 Model | Calibration | 0.403 | 1.466 | 0.45 | 404 | 468 | 0.074 | 2.09 | <0.001 | 2047 |
| | Evaluation | −0.234 | 1.116 | 0.49 | 224 | 185 | 0.082 | 1.98 | <0.001 | 800 |

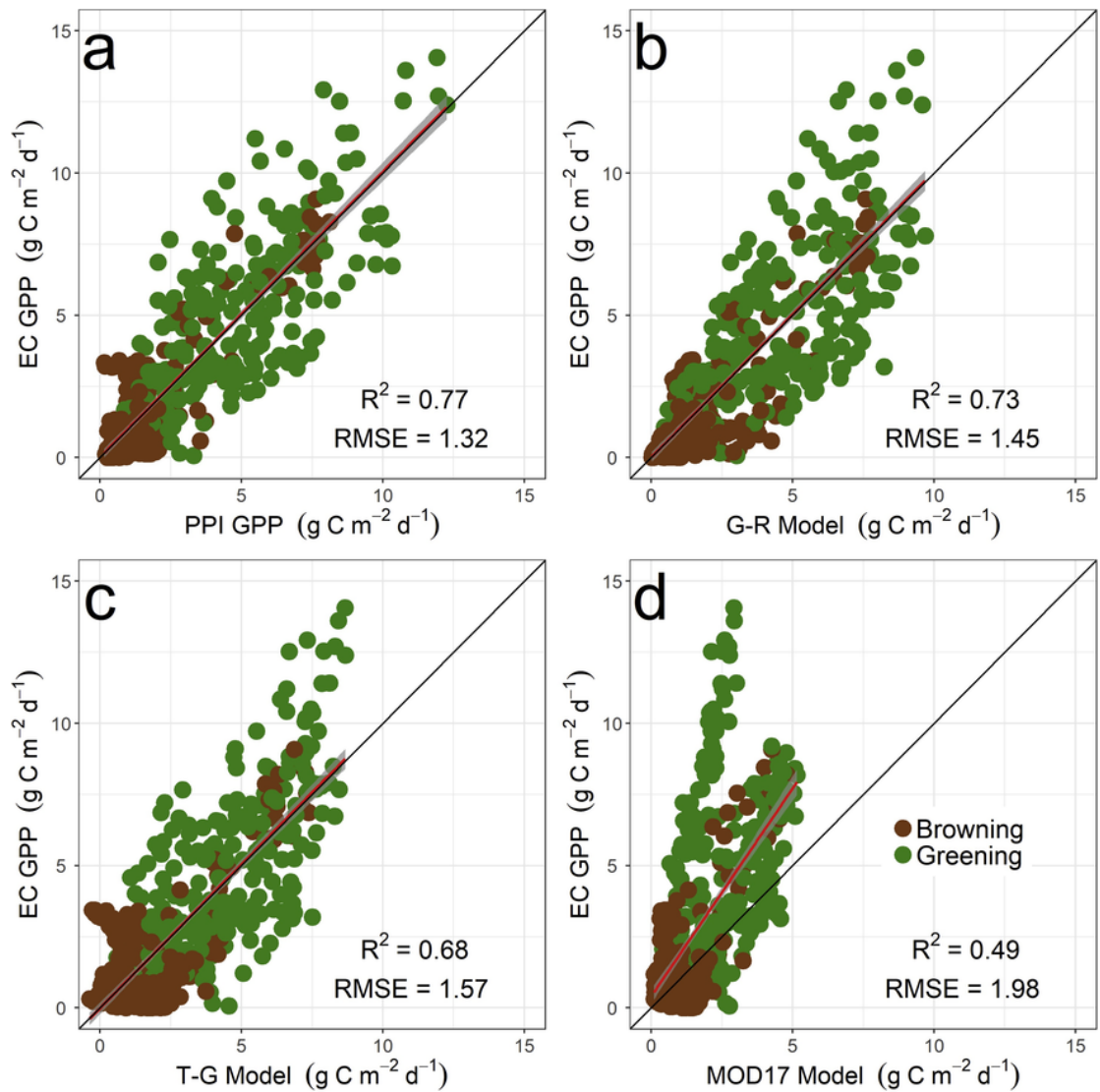

**Fig. 7.** Site-level relationships between EC GPP and each of the four GPP models based on the evaluation data (a – d, see Table 4 for the statistical summary). The solid black line is the 1:1 line, the solid red line is the regression line, and the gray shading is the 95% confidence interval.

### 3.2. Relationship between field and EO data

Comparison between EC GPP, PPI, LST, VPD and SWC (soil water content) are shown in Fig. 3. Overall, there is a discernible limitation imposed by VPD on EC GPP at 20 hPa. This pattern was clear for Skukuza (30% TCC, MAT = 22 °C) and Mongu (65% TCC, MAT = 25 °C) where most of the assimilation took place at VPD < 20 hPa. However, VPD limitation at Dahra (3% TCC, MAT = 29 °C) and Demokeya (7% TCC, MAT = 30 °C) was less clear as both sites showed continued assimilation past 20 hPa albeit at a much lower magnitude. This could be due to differences in mean annual VPD (MVPD) between the Sahelian and southern African sites. Both Dahra (MVPD = 23.8 hPa) and Demokeya (MVPD = 26.2 hPa) have higher atmospheric moisture demand throughout the year than Skukuza (MVPD = 11.4 hPa) and

Mongu (MVPD = 12.7 hPa). The low assimilation (< 5 g C m2 d$^{-1}$) at VPD > 20 hPa in Fig. 3 c could indicate that the tree species that dominate the sparse canopies of warmer sites have developed adaptations that can withstand higher VPD (Klein et al., 2014; Smith et al., 2004). These adaptations may include reduced metabolic activity that imposes a reduction on carbon assimilation and access to deeper water (Ardö et al., 2008; Shirke and Pathre, 2004).

The SWC constraint on EC GPP was clear below 5% when assimilation either sharply declines or ceases altogether (Fig. 3 d). The exception is Mongu, which continued to assimilate a maximum of 7.7 g C m2 d$^{-1}$ at 1.3% SWC, whereas Dahra practically ceases assimilation below 2% SWC. The wilting point of the upper layers of the soil at these sites is around 2–5%, and sustained assimilation so close to that point can be explained by the trees having access to deeper water (Abdi et al., 2017; Ardö et al., 2008). Nevertheless, most of the carbon assimilation across the four sites takes place above 5% SWC.

Hysteresis, a disproportionate pattern that creates the appearance of a loop in the scatterplot, is observed in the relationship between EC GPP and PPI (Fig. 3 a). This could be caused by variation in canopy chlorophyll content over the growing season, particularly during the greening and browning phases (Gitelson et al., 2014). Fig. 4 shows this in detail for Mongu. There is a strong link between EC GPP and PPI during the greening phase as the soil moisture is replenished and VPD drops. The start of the browning phase is evidenced by decrease in soil moisture levels and an increase in VPD. Leaf senescence during this phase is a gradual process and plants undergo slower change of canopy chlorophyll relative to greening phase due to the process of leaf aging (Jenkins et al., 2007; Muraoka et al., 2013; Richardson et al., 2013). Canopy chlorophyll is linearly related to PPI during the greening phase as indicated in the tight coupling of EC GPP and PPI (Fig. 4), but the relationship turns non-linear during the browning phase. This indicates that the contribution of canopy chlorophyll to GPP differs between phases, and Gitelson et al. (2014) suggested that it is possibly due to the vertical distribution of green LAI and chlorophyll content during the growing season. Another reason for this hysteresis is the combined effect of SWC and VPD during the greening and browning phases. During the greening phase, SWC is replenished, VPD decreases and the stomata fully opens to take up carbon. During the browning phase, SWC declines and VPD increases with increasing temperature causing the stomata gradually close in order to conserve water (Abdi et al., 2017; Pingintha et al., 2010).

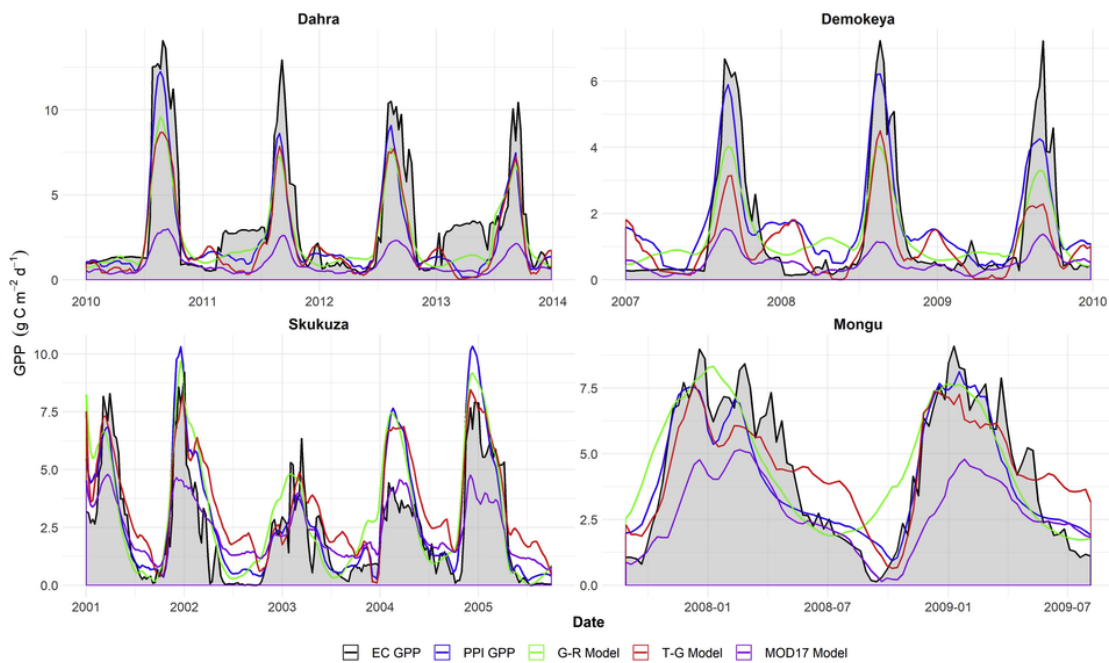

**Fig. 8.** Time series comparison between the four GPP models (see Table 3 for model statistics) at each site: **(a)** Demokeya, **(b)** Dahra, **(c)** Skukuza, and **(d)** Mongu. The gray shading indicates the area under EC GPP.

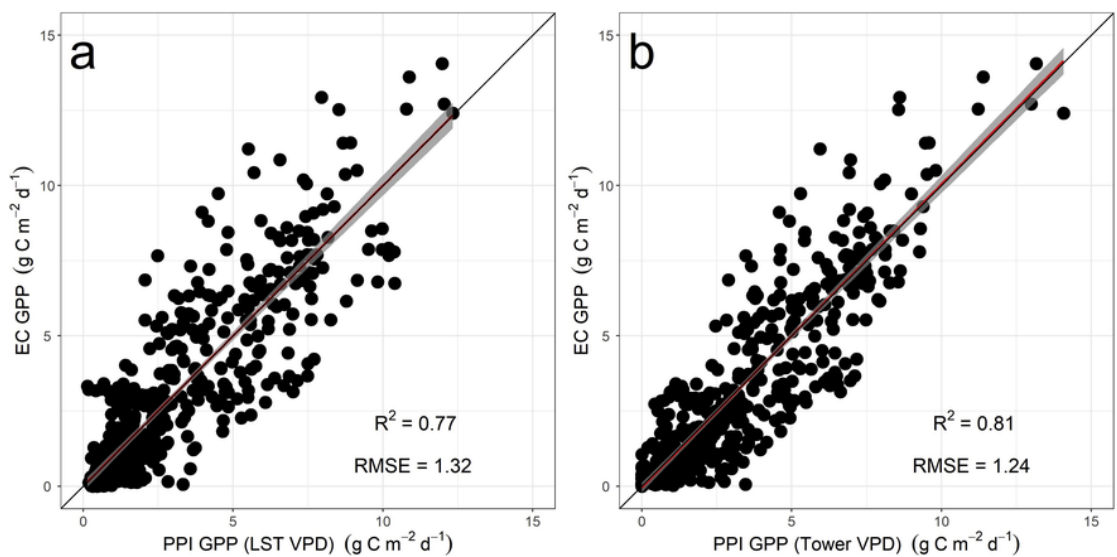

**Fig. 9.** Comparison of two PPI GPP models, one using scaled VPD derived from MODIS LST (a) and one using eddy covariance flux tower VPD (b). The solid black line is the 1:1 line, and the gray shading is the 95% confidence interval.

The relationship between EC GPP and LST exhibited a triangular feature space that slopes towards lower temperatures with increasing EC GPP (Fig. 3 b). Approximately half of the total carbon uptake across all four sites took place between 30 °C and 40 °C. There was a 40% reduction in EC GPP at LST $\geq$ 45 °C (1.13 ± 1.20 g C m$^{-2}$ d$^{-1}$) relative to LST $<$ 45 °C (2.86 ± 2.81 g C m$^{-2}$ d$^{-1}$). VPD and LST have a generally strong relationship across all the sites (R$^2$ = 0.67) (Fig. 5), due to the connection of LST with surface moisture conditions and the partitioning of latent and sensible heat fluxes. Generally, the rainy season in these ecosystems has a lower LST than the dry season due to increased vegetative cover, and the incoming energy is rapidly utilized by evapotranspiration (Nutini et al., 2014).

The parameterization summary of PPI is show in Table 3 and the bootstrap distribution histogram of the slope is shown in Fig. 6. Overall, PPI correlated strongly with EC GPP ($R^2 = 0.77$; RMSE = 1.45 g C $m^2$ $d^{-1}$; BIC = 1707) in the calibration phase. Similarly, in the evaluation phase PPI exhibited a strong correlation with EC GPP ($R^2 = 0.75$; RMSE = 1.41 g C $m^2$ $d^{-1}$; BIC = 673), indicating that the spatiotemporal variability in EC GPP can be captured by PPI across the four sites. The linear regression model for parametrizing PPI is:

$$PPI_{parameterized} = 6.166 \times PPI + 0.516 \tag{13}$$

### 3.3. Comparison of GPP models

The PPI GPP model demonstrated an overall better performance in predicting EC GPP than the three other models (Table 4). In the evaluation, the explained variance was highest for PPI GPP ($R^2 = 0.77$, Fig. 7a) followed by the G–R model ($R^2 = 0.73$, Fig. 7b) and T-G model ($R^2 = 0.68$, Fig. 7c). The MOD17 model had the lowest accuracy ($R^2 = 0.49$) and underestimated EC GPP. The MOD17 model displayed two distinct clusters, a higher slope at $\beta_1 = 4.5$ and a lower one at $\beta_1 = 1.74$ (Fig. 7d). The data points within the higher slope belong to the two Sahelian sites, and the lower slope includes the two southern African sites. This split could be attributed to MOD17's reliance on the MOD12Q1 land cover product to populate the biome properties look-up table (BPLUT) that serves as the basis for establishing optimum LUE for each biome type (Friedl et al., 2010; Zhao et al., 2011). MOD-12Q1 classifies Mongu as "Savanna", a class that has a maximum possible tree canopy cover of 30% according to Hansen et al. (2000), which is roughly half of the actual canopy cover at the site. Indeed, classification accuracies of MOD12Q1 for the land cover types that dominate our four sites are rather poor. The producer's and user's accuracies of MOD12Q1 for woody savanna were 45.2 ± 4.1% and 34.3 ± 4.5%, respectively, and for savanna they are 22.6 ± 4.4% and 39 ± 6%, respectively (Friedl et al., 2010).

A time series comparison of the four GPP models (PPI, G–R, T-G, and MOD17) and EC GPP for each site is shown in Fig. 8. The widely-used MOD17 model underestimated GPP at all sites but was able to consistently follow the greening and browning phases of Skukuza (30% canopy cover) and Mongu (65% canopy cover), suggesting that this model does better in tree-dominated areas than savanna. The T-G model either underestimated or overestimated peak GPP at most sites. The PPI GPP model tracked the seasonal and inter-annual development of EC GPP and captured its amplitude reasonably well. At Demokeya, the PPI model over-estimated GPP during the dry season, but captured the start and end of the growing season for all sites in-line with EC GPP except at Mongu. The final PPI GPP model used is shown in Eq. 14. This model is based on Eq. 4 and it accounts for the observed relationship between EC GPP and VPD (Fig. 3c).

$$PPI\_GPP = \left( \overbrace{6.166 \times PPI + 0.516}^{PPI_{parameterized}} \right) \times \left( VPD_{Scaled} \right)^{\frac{2}{3}} \tag{14}$$

### 3.4. VPD as a down-regulation scalar instead of LST

The configuration of the T-G model should theoretically account for water stress because of the direct inclusion of LST. The model has been found to perform poorly in water-limited woody savanna, savanna and grasslands, and the application of a Markov Chain Monte Carlo optimization by Dong et al. (2017) did not improve its performance in those ecosystems. The PPI GPP model outperformed the T-G model in both the calibration and evaluation datasets (Table 4). The sites classified as "drought sites" in Sims et al. (2008) are considerably cooler than Sahelian sites (12 °C vs 30 °C), so the LST scaling factor in the T-G model cannot account for the environmental conditions in this warm region. However, the T-G model performed well at Skukuza ($R^2 = 0.73$) and Mongu ($R^2 = 0.79$), probably because these sites have a lower mean annual temperatures (29 °C and 25 °C, respectively) that is within the threshold set in the T-G model (Sims et al., 2008).

Substituting $VPD_{LST}$ with scaled site-based $VPD_{TOWER}$ improves EC GPP variance that is explained by the model (Fig. 9). An explanation for the reduced performance of $VPD_{LST}$ relative to $VPD_{TOWER}$ is that the uncertainty in $VPD_{LST}$, which is a function of the relationship between satellite-derived LST and VPD (Eq. 2), is disseminated in the PPI GPP model as per the law of error propagation. This is not surprising considering the fact that $VPD_{LST}$ has an average RMSE of 4.31 hPa across all four sites (Fig. 5). This highlights the tradeoff between model performance and upscaling land surface processes (Leitão et al., 2018).

Generally, the inclusion of LST allows for a GPP model to capture heat stress, however the direct effect of LST (i.e. actual canopy temperature) on photosynthesis is rather unclear (Wu et al., 2010). The mechanism through which high canopy temperatures affect photosynthesis is not well understood. There is evidence that decrease in photosynthesis at high canopy temperatures is caused by high leaf-to-air VPD than by direct negative effects of temperature on photosynthetic metabolism (Lloyd and Farquhar, 2008). Indeed, the physiological response of vegetation is strongly linked to the terrestrial surface energy balance (Chapin et al., 2011; Duveiller et al., 2018). There is also some indication that increasing LST over a vegetation canopy is linked to looming drought due to decrease in latent heat flux (stomatal closure) and increase in sensible heat flux (Wan et al., 2004).

According to Fick's first law of diffusion, plants lose water to the atmosphere with increasing VPD if the stomata remain open (Farquhar and Sharkey, 1982). Therefore, it is reasonable to assume that there will be a decrease in transpiration with rising VPD because plants will opt to conserve water. This means that for each of our four sites, there will be a reduction in the amount of energy leaving the canopy in the form of latent heat. Because energy fluxes must balance, there will be a corresponding increase in sensible heat that raises leaf temperature. It is plausible that this increase in sensible heat manifests as increase in LST. As leaf temperature increases, a weakening of the biochemical processes that occur during photosynthesis and can further reduce GPP. Thus, plant physiological response to increased VPD may trigger a process that is captured in the remotely-sensed LST signal.

## 4. Conclusions

The terrestrial carbon flux drives several ecosystem functions and estimating it at large scales is a necessary undertaking in global environmental change research. The plant phenology index (PPI) was originally designed for boreal coniferous forests as a solution to suppressing the influence of snow in phenology metrics, however, due to its general formulation it works in other environments as well. In this study, we developed and evaluated the performance of a PPI-based model in predicting the gross primary productivity (GPP) at four semi-arid sites in sub-Saharan Africa with a wide canopy cover range (3–65%). We found that the PPI GPP model captured the magnitude of eddy covariance flux tower-measured GPP relatively well compared to the other tested models ($R^2 = 0.77$ and RMSE = 1.32 g C $m^{-2}$ $d^{-1}$) due to its sensitivity to green LAI, and therefore canopy chlorophyll content. The better performance of the PPI GPP model is also due to the inclusion of a water stress variable (VPD) as a down-regulating scalar. The greenness and radiation (G–R) model does not have a direct water stress scalar, and the temperature and greenness (T-G) model uses LST as an indirect substitute to water stress that does not capture the direct physiological response to that stress. The MOD17 production efficiency model includes VPD as a down-regulating scalar, however the model is dependent on the MODIS land cover product for its look-up table parameter allocation, which performs poorly in savanna and woody savanna biomes. Further research is needed to explore the strong link between PPI and green LAI, and develop robust models for semi-arid ecosystems using PPI as an indicator of photosynthetic capacity.

## Author contributions

AMA conceived and designed the study, conducted the analysis, interpreted the results, and drafted the manuscript. HJ, LE, and JA assisted with the study design. NBO and TT contributed to the interpretation of the results. NBO and VL processed the PPI data and approved the final draft. All authors extensively discussed the results and revised the manuscript.

## Acknowledgments

AMA acknowledges support from the Lund University Center for Studies of Carbon Cycle and Climate Interactions (LUCCI) and the Royal Physiographic Society in Lund. Additional funding for this project was provided by the Swedish National Space Board (contract no. 100/11 to JA) and by the Department of Physical Geography and Ecosystem Science of Lund University. The authors would like to thank the reviewers for their time in providing detailed and constructive comments that improved the quality of the manuscript.